\documentclass[%
  reprint,
  showpacs,preprintnumbers,
  amsmath,amssymb,
  aps,
  prb,
uplatex
]{revtex4-2}

\usepackage{graphicx}[]
\usepackage{dcolumn}
\usepackage{bm}
\usepackage{multirow}
\usepackage{braket}
\usepackage{color}
\usepackage{ulem}

\usepackage{hyperref}
\usepackage{url}

\begin{document}

\preprint{APS/123-QED}

\title{
Locking of skyrmion cores on a centrosymmetric discrete lattice: onsite versus offsite
}

\author{Satoru Hayami and Ryota Yambe}
\affiliation{
Department of Applied Physics, The University of Tokyo, Tokyo 113-8656, Japan
}
 
\begin{abstract}
A magnetic skyrmion crystal (SkX) with a swirling spin configuration, which is one of topological spin crystals as a consequence of an interference between multiple spin density waves, shows a variety of noncoplanar spin patterns depending on a way of superposing the  waves. 
By focusing on a phase degree of freedom among the constituent waves in the SkX, we theoretically investigate a position of the skyrmion core on a discrete lattice, which is relevant with the symmetry of the SkX.
The results are obtained for the double exchange (classical Kondo lattice) model on a discrete triangular lattice by the variational calculations. 
We find that the skyrmion cores in both two SkXs with the skyrmion number of one and two are locked at the interstitial site on the triangular lattice, while it is located at the onsite by introducing a relatively large easy-axis single-ion anisotropy. 
The variational parameters and the resultant Fermi surfaces in each SkX spin texture are also discussed. 
The different symmetry of the Fermi surfaces depending on the core position is obtained when the skyrmion crystal is commensurate with the lattice. The different Fermi-surface topology is directly distinguished by an electric probe of angle-resolved photoemission spectroscopy. 
Furthermore, we show that the SkXs obtained by the variational calculations are also confirmed by numerical simulations on the basis of the kernel polynomial method and the Langevin dynamics for the double exchange model and the simulated annealing for an effective spin model. 
\end{abstract}
\maketitle

\section{Introduction}
\label{sec:Introduction}

A wave nature of electrons provides millions of enigmatic problems since the early twentieth century.  
In particular, an interference among electron wave functions in solids gives rise to unconventional electronic structures and transport properties, which makes considerable advances of modern science technology.  
For example, specific atomic alignments on lattices bring about unconventional electronic states with massless Dirac fermions, flat-band structures, and quasi-crystal nature as recently observed in the twisted bilayer graphene~\cite{bistritzer2011moire,cao2018unconventional,cao2018correlated,Koshino_PhysRevX.8.031087,yao2018quasicrystalline}. 
Another example is found in quantum interference effects through electron dynamics, such as the Anderson localization in disordered systems~\cite{Anderson_PhysRev.109.1492,Abrahams_PhysRevLett.42.673}, anomalous quantum Hall effect in magnetic topological insulators~\cite{Haldane_PhysRevLett.61.2015,chang2013experimental}, and Josephson effect in superconductors~\cite{josephson1962possible,Ambegaokar_PhysRevLett.10.486}.

The interference also occurs in magnetic materials by superposing spin density waves dubbed the multiple-$Q$ magnetic state. 
An important consequence of the multiple-$Q$ formation is to induce a superstructure of spin and scalar spin chirality, the latter of which is defined as a scalar triple product of spins, $\chi_{\bm{R}}=\bm{S}_i \cdot (\bm{S}_j \times \bm{S}_k)$.
In particular, the crystallization of $\chi_{\bm{R}}$ can lead to a quantized topological (skyrmion) number $n_{\rm sk}$, which is defined by~\cite{BERG1981412} 
\begin{align}
\label{eq:nsk_num}
n_{\rm sk}=\frac{1}{4\pi N_m}\sum_{i}  2\tan^{-1} \frac{\bm{S}_i \cdot (\bm{S}_j \times \bm{S}_k)}{1+\bm{S}_i \cdot \bm{S}_j+\bm{S}_j \cdot \bm{S}_k+\bm{S}_k \cdot \bm{S}_i},
\end{align} 
where $N_m$ is the number of magnetic unit cell in the discrete lattice system. 
An archetype of multiple-$Q$ magnetic states with a nonzero integer topological number is a skyrmion crystal (SkX), which has been widely studied in $d$- and $f$-electron compounds~\cite{Bogdanov89,Bogdanov94,nagaosa2013topological,rossler2006spontaneous,Muhlbauer_2009skyrmion,yu2010real,seki2012observation,kezsmarki_neel-type_2015,kakihana2018giant,kaneko2019unique,tabata2019magnetic,kakihana2019unique,hayami2021field}.  
The experimental findings mainly correspond to the SkX with $|n_{\rm sk}|=1$, which is characterized by a superposition of multiple cycloidal or proper-screw spirals, while it is possible to construct
other types of the SkXs, i.e. the SkX with $|n_{\rm sk}|=2$, by adopting a superposition of multiple sinusoidal waves from the theoretical viewpoint~\cite{Ozawa_PhysRevLett.118.147205,Hayami_PhysRevB.99.094420,amoroso2020spontaneous,yambe2021skyrmion,amoroso2021tuning}.  
Furthermore, the interference between spin density waves gives rise to various multiple-$Q$ spin textures depending on the types of constituent waves and a relative phase between them: 
a hedgehog lattice~\cite{tanigaki2015real,kanazawa2017noncentrosymmetric,fujishiro2019topological,Ishiwata_PhysRevB.84.054427,Ishiwata_PhysRevB.101.134406,Rogge_PhysRevMaterials.3.084404,Binz_PhysRevB.74.214408,Park_PhysRevB.83.184406,Okumura_PhysRevB.101.144416,grytsiuk2020topological}, meron crystal~\cite{yu2018transformation,kurumaji2019skyrmion,Hayami_PhysRevLett.121.137202,Hayami_PhysRevB.103.024439,Wang_PhysRevB.103.104408,Utesov_PhysRevB.103.064414,Hayami_PhysRevB.104.094425}, bubble crystal~\cite{lin1973bubble,Garel_PhysRevB.26.325,takao1983study,Hayami_PhysRevB.93.184413,seo2021spin}, vortex crystal~\cite{Momoi_PhysRevLett.79.2081,Kamiya_PhysRevX.4.011023,Wang_PhysRevLett.115.107201,Marmorini2014,Hayami_PhysRevB.94.174420,takagi2018multiple,hayami2020phase}, and chirality density wave~\cite{Solenov_PhysRevLett.108.096403,Ozawa_doi:10.7566/JPSJ.85.103703,Hayami_PhysRevB.94.024424,Shimokawa_PhysRevB.100.224404,yambe2020double}.

From an energetic point of view, the emergence of these multiple-$Q$ states depends on the model Hamiltonian and the lattice structure~\cite{hayami2021topological}. 
For example, the types of constituent waves in the SkXs are determined by microscopic interactions; the Heisenberg model with the ferromagnetic interaction and the Dzyaloshinskii-Moriya (DM) interaction in chiral (polar) magnets tends to favor the Bloch (N\'eel) SkX consisting of proper-screw (cycloidal) spirals, while the double exchange (classical Kondo lattice) model on a centrosymmetric square lattice hosts either the Bloch, N\'eel, or anti-SkX depending on the signs of the anisotropic Ruderman-Kittel-Kasuya-Yosida (RKKY) interactions~\cite{Ruderman,Kasuya,Yosida1957} in momentum space~\cite{Hayami_doi:10.7566/JPSJ.89.103702}. 
Similarly, it is recognized that the phase degree of freedom in the multiple-$Q$ states is closely related to interactions, which brings about the phase transitions between the different types of multiple-$Q$ states~\cite{hayami2020phase,Shimizu_PhysRevB.103.054427,Shimizu_PhysRevB.103.184421}. 
The long-range chirality interaction leads to a topological transition between the SkX with $|n_{\rm sk}|=2$ and the tetra-axial vortex crystal with $|n_{\rm sk}|=0$ by changing the relative phase among the constituent waves~\cite{hayami2020phase}, and magnetic anisotropy arising from the mirror symmetry breaking drives a transition between the SkXs with $|n_{\rm sk}|=1$ and $|n_{\rm sk}|=2$ by changing the relative phase between $xy$ and $z$ components of the individual spiral~\cite{yambe2021skyrmion}. 
These studies suggest that there is a further possibility to induce a new type of multiple-$Q$ states by considering specific model parameters and lattice structures, which might be useful to stimulate further experimental findings of exotic topological spin crystals beyond the SkX and the hedgehog lattice~\cite{gobel2021beyond,hayami2021topological}. 
It is also desirable to provide a way to identify different multiple-$Q$ states experimentally, since it is sometimes difficult to distinguish them by diffraction techniques like the neutron scattering and the resonant x-ray scattering.

In the present study, we investigate the formation of the SkXs in itinerant magnets on a discrete lattice by focusing on a position of the skyrmion core, which is related to the phase degree of freedom in the SkXs. 
We consider the short-period SkXs in the double exchange (classical Kondo lattice) model with an easy-axis single-ion anisotropy on a triangular lattice, where the SkXs with $|n_{\rm sk}|=1$ and $|n_{\rm sk}|=2$ consisting of the commensurate ordering wave vectors with the crystal lattice are stabilized with and without the magnetic field~\cite{Ozawa_PhysRevLett.118.147205}. 
By constructing a phase diagram within the variational calculations, we find that the position of the skyrmion core is located at the interstitial site for a small single-ion anisotropy, while that is locked at the onsite for a large single-ion anisotropy. 
Reflecting the different core positions, the resultant electronic states for the SkXs with $|n_{\rm sk}|=1$ are sensitively affected in a different way: 
the threefold-symmetric Fermi surface appears for a small anisotropy, while the sixfold-symmetric one appears for a large anisotropy. 
Furthermore, we show that the SkX with $|n_{\rm sk}|=1$ is described by superposing of three spiral and sinusoidal waves for a small anisotropy, which is in contrast to the superposition of three spirals for the SkX in chiral/polar magnets. 
We confirm that a part of the SkXs obtained by the variational calculations are reproduced by the unbiased kernel polynomial method and the Langevin dynamics (KPM-LD) simulations for the large system size~\cite{Barros_PhysRevB.88.235101}. 
We also perform the simulated annealing for an effective spin model of the double exchange model to discuss the SkX spin texture for a large single-ion anisotropy.

The rest of the paper is organized as follows. 
After presenting the double exchange model in Sec.~\ref{sec:Kondo lattice model}, we show the results by the variational calculations for the double exchange model on the triangular lattice in Sec.~\ref{sec:Variational calculations}. 
We discuss the optimal variational parameters and the Fermi surfaces while changing the magnetic field and easy-axis single-ion anisotropy. 
We examine the stability of the SkXs by using the unbiased KPM-LD simulations in Sec.~\ref{sec:Langevin dynamics simulation}. 
We also show the result by the simulated annealing for the effective spin model. 
We summarize our results in Sec.~\ref{sec:Summary}.

\section{Double exchange model}
\label{sec:Kondo lattice model}

We consider an itinerant electron model consisting of noninteracting electrons coupled with localized spins, dubbed the double exchange (classical Kondo lattice) model, on the triangular lattice under the periodic boundary condition. 
The Hamiltonian is given by 
\begin{align}
\label{eq:Ham}
\mathcal{H} = &-t_1 \sum_{\langle i, j \rangle \sigma} (c^{\dagger}_{i\sigma}c_{j \sigma}+ {\rm H.c.})
-t_3 \sum_{\langle \langle i, j \rangle \rangle \sigma} (c^{\dagger}_{i\sigma}c_{j \sigma}+ {\rm H.c.}) \nonumber \\
&+J \sum_{i} \bm{s}_i
\cdot \bm{S}_i -\mu\sum_{i,\sigma}c^{\dagger}_{i\sigma}c_{i\sigma}
-A \sum_{i} (S_i^z)^2
-H \sum_{i} S_i^z, 
\end{align}
where $c^{\dagger}_{i\sigma}$ ($c_{i \sigma}$) is a creation (annihilation) operator of an itinerant electron at site $i$ with spin $\sigma$. 
The first and second terms represent the kinetic energies of itinerant electrons, where the sums of $\langle i,j \rangle$ and $\langle \langle  i,j \rangle \rangle$ are taken over the nearest- and third-neighbor sites on the triangular lattice, respectively. 
The third term represents the exchange coupling between itinerant electron spins $\bm{s}_i = (1/2)\sum_{\sigma \sigma'}c^{\dagger}_{i\sigma} \bm{\sigma}_{\sigma \sigma'} c_{i \sigma'}$ and localized spins $\bm{S}_i$, where $\bm{\sigma}=(\sigma^x,\sigma^y,\sigma^z)$ is the vector of Pauli matrices and $\bm{S}_i$ is regarded as a classical spin whose amplitude is normalized as $|\bm{S}_i|=1$. 
The sign of $J$ is irrelevant for classical localized spins. 
The fourth term represents the chemical potential $\mu$. 
The fifth term represents the easy-axis single-ion anisotropy ($A>0$).
The sixth term represents the Zeeman coupling to an external magnetic field along the $z$ direction.
We consider the effect of the magnetic anisotropy and the magnetic field only on localized spins for simplicity. 
Hereafter, we take $t_1=1$ as the energy unit of the double exchange model. 
We also set the lattice constant $a=1$ as the length unit of the triangular lattice. 

The other model parameters are chosen so that the SkXs are stabilized in the ground state and their period is commensurate with that of the lattice: $t_3=-0.85$, $J=1$, and $\mu=-3.5$~\cite{Ozawa_PhysRevLett.118.147205}. 
For these parameters, the bare susceptibility of itinerant electrons shows a triple peak structure in momentum space at $\bm{Q}_1=(Q,0)$, $\bm{Q}_2=(-Q/2,\sqrt{3}Q/2)$, and $\bm{Q}_3=(-Q/2,-\sqrt{3}Q/2)$ with the relatively large ordering vector $Q=\pi/3$. 
The three ordering vectors, $\bm{Q}_1$, $\bm{Q}_2$, and $\bm{Q}_3$, are connected by the threefold rotational symmetry. 
Consequently, the short period triple-$Q$ states are stabilized in the ground state: the $|n_{\rm sk}|=2$ SkX for $0 \leq H \lesssim 0.00325$, the $|n_{\rm sk}|=1$ SkX for $0.00325 \lesssim H \lesssim 0.0065$, and another triple-$Q$ state with $|n_{\rm sk}|=0$ for $0.0065 \lesssim H$ in the absence of $A$~\cite{Ozawa_PhysRevLett.118.147205}. 
By introducing the easy-axis anisotropy, the stable region for the $|n_{\rm sk}|=1$ SkX is extended, e.g., $0.004 \lesssim H \lesssim 0.0115$ for $A=0.002$~\cite{Hayami_PhysRevB.99.094420}. 
Meanwhile, the detailed multiple-$Q$ nature of the SkXs, especially for the $|n_{\rm sk}|=1$ SkX, was not fully shown, as described below.  
In the following, we discuss the details of the SkXs by explicitly taking into account the phase degree of freedom among the constituent waves with the use of the variational calculations.

\section{Variational calculations}
\label{sec:Variational calculations}

We show the result by the variational calculation in this section. 
In order to discuss the optimal spin configurations of the SkXs and their modulations while varying the model parameters $A$ and $H$ in Eq.~(\ref{eq:Ham}), we consider a general form to represent the triple-$Q$ structures, which is constructed by the superposition of the three spin density waves with $\bm{Q}_1$, $\bm{Q}_2$, and $\bm{Q}_3$ as 
\begin{align}
\label{eq:spin}
\bm{S}_i  \propto 
\left(
\begin{array}{c}
\sin \mathcal{Q}'_{1i} - \frac{1}{2} \sin \mathcal{Q}'_{2i} - \frac{1}{2} \sin \mathcal{Q}'_{3i} \\
 \frac{\sqrt{3}}{2} \sin \mathcal{Q}'_{2i} - \frac{\sqrt{3}}{2} \sin \mathcal{Q}'_{3i} \\
 \tilde{M}_z- a_z (\cos \mathcal{Q}_{1i}+ \cos \mathcal{Q}_{2i}+\cos \mathcal{Q}_{3i})
\end{array}
\right)^{\rm T}, 
\end{align}
where ${\rm T}$ denotes the transpose of the vector. 
In the spin configuration, $\mathcal{Q}_{\nu i}=\bm{Q}_\nu \cdot \bm{r}_i + \theta_\nu$ ($\nu=1$-$3$ and $\bm{r}_i$ is the position vector at site $i$) and $\mathcal{Q}'_{\nu i}=\mathcal{Q}_{\nu i}+ \psi$; $\theta_\nu$, $\psi$, $a_z$, and $\tilde{M}_z$ are the variational parameters. 
We choose the specific vorticity and helicity, which are defined by the winding number projected onto the $S^x$-$S^y$ plane around the skyrmion core and the relative angle in the $S^x$-$S^y$ plane, respectively~\cite{nagaosa2013topological}, because they are arbitrary owing to rotational symmetry around the $z$ axis in spin space in the double exchange model in Eq.~(\ref{eq:Ham}). 
We do not consider the anisotropic triple-$Q$ state with different amplitudes at $\bm{Q}_1$-$\bm{Q}_3$, which is stabilized for $0.006 \lesssim H$ at $A=0$~\cite{Ozawa_PhysRevLett.118.147205}, in order to focus on the optimal spin configurations of the SkXs described by Eq.~(\ref{eq:spin}) for simplicity.

\begin{figure}[t!]
\begin{center}
\includegraphics[width=1.0 \hsize ]{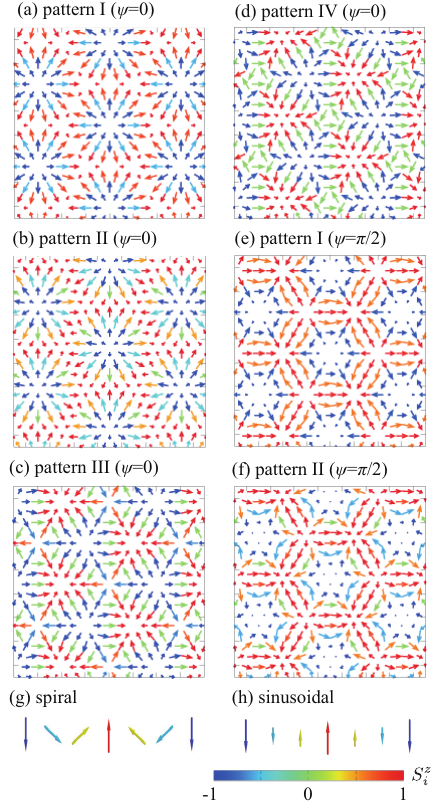} 
\caption{
\label{fig:PD}
(a)-(f) Schematics of the variational spin textures in (a) pattern I with $\psi=0$, (b) pattern II with $\psi=0$, (c) pattern III with $\psi=0$, (d) pattern IV with $\psi=0$, (e) pattern I with $\psi=\pi/2$, and (f) pattern II with $\psi=\pi/2$. 
The arrows represent the direction of the in-plane spin moment and the color shows its $z$ component. 
The empty sites have the spin moment with $S_i^z=\pm 1$. 
(g), (h) Schematics of (g) the spiral waves and (h) the sinusoidal waves. 
The spin textures in (a)-(d) consists of (g), while those in (e) and (f) consists of (h).} 
\end{center}
\end{figure}

\begin{figure}[t!]
\begin{center}
\includegraphics[width=1.0 \hsize]{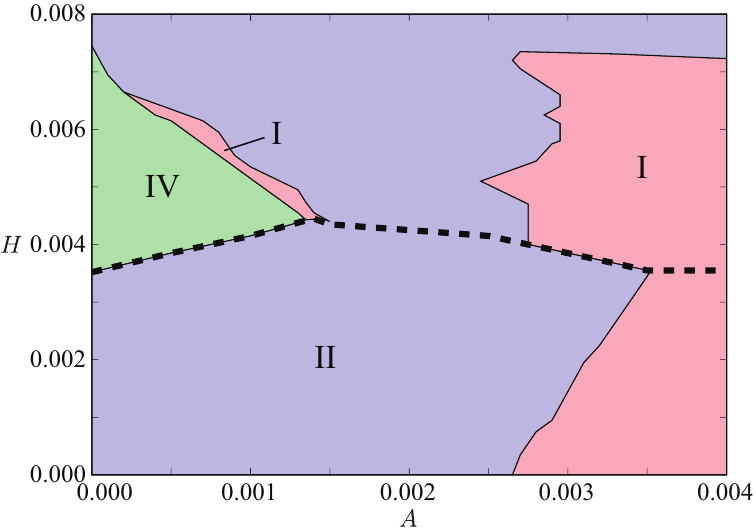} 
\caption{
\label{fig:PD12}
The ground-state phase diagram while changing $A$ and $H$ obtained by the variational calculations for the system size $N=96^2$. 
The dashed curve represents the boundary between the $|n_{\rm sk}|=1$ and $|n_{\rm sk}|=2$ SkXs. 
}
\end{center}
\end{figure}

There are two phase degrees of freedom in Eq.~(\ref{eq:spin}). 
The one is $\theta_\nu$, which represents the relative phase degree of freedom among the constituent waves.  
We here consider four sets of $(\theta_1,\theta_2,\theta_3)$ in the variational calculations: (0,0,0) denoted as the pattern I, $(\pi/3,-\pi/3,0)$ denoted as the pattern II, $(\pi/2,\pi/2,\pi/2)$ denoted as the pattern III, and $(5\pi/6,\pi/6,\pi/2)$ denoted as the pattern IV. 
The patterns I and II with $\sum_\nu \theta_\nu=0$ represent the SkX spin textures and the patterns III and IV with $\sum_\nu \theta_\nu=3\pi/2$ represent the meron-antimeron crystal spin textures, as shown in Figs.~\ref{fig:PD}(a)-\ref{fig:PD}(d). 
The main difference between them is found in the net scalar chirality at zero field ($\tilde{M}_z =0$). 
The SkX spin texture exhibits a nonzero net scalar chirality $\chi^{\rm sc}=\sum_{\bm{R}}\chi_{\bm{R}}$, where $\bm{R}$ is the position vector for the center of the triangle plaquette, while the meron-antimeron crystal one does not have $\chi^{\rm sc}$ owing to the cancellation of positive and negative $\chi_{\bm{R}}$. 
It is noted that the patterns III and IV carry a nonzero net scalar chirality for $\tilde{M}_z
\neq 0$ in the external magnetic field, which can also have a quantized skyrmion number~\cite{Hayami_PhysRevB.104.094425}. 

Meanwhile, the difference between the patterns I and II (or patterns III and IV) is found in the position of the skyrmion core defined as the position with $S^z_i=-1$. 
The skyrmion cores in the patterns I and III are lied at the sites, while those in the patterns II and IV are lied at the interstitial sites, as shown in Figs.~\ref{fig:PD}(a)-\ref{fig:PD}(d).
In other words, the constitute waves with the same phase ($\theta_1=\theta_2=\theta_3$) fix the core at the sites, while those with different phases remove it from the sites. 

The other phase degree of freedom in the SkXs is the relative phase between the $xy$- and $z$-spin components, $\psi$, which has been explicitly pointed out in Ref.~\onlinecite{yambe2021skyrmion}. 
This phase degree of freedom is related to the nature of the spin density waves for individual wave vectors.
The state with $\psi=0$ represents the spiral wave [Fig.~\ref{fig:PD}(g)], while that with $\psi=\pi/2$ represents the sinusoidal wave [Fig.~\ref{fig:PD}(g)] along each $\bm{Q}_\nu$ direction. 
Reflecting the different constituent waves, the triple-$Q$ SkX states show different skyrmion numbers. 
For example, the patterns I and II with $\psi=0$ have $|n_{\rm sk}|=1$ [Figs.~\ref{fig:PD}(a) and \ref{fig:PD}(b)], while those with $\psi=\pi/2$ have $|n_{\rm sk}|=2$ [Figs.~\ref{fig:PD}(e) and \ref{fig:PD}(f)].

\begin{figure*}[t!]
\begin{center}
\includegraphics[width=1.0\hsize]{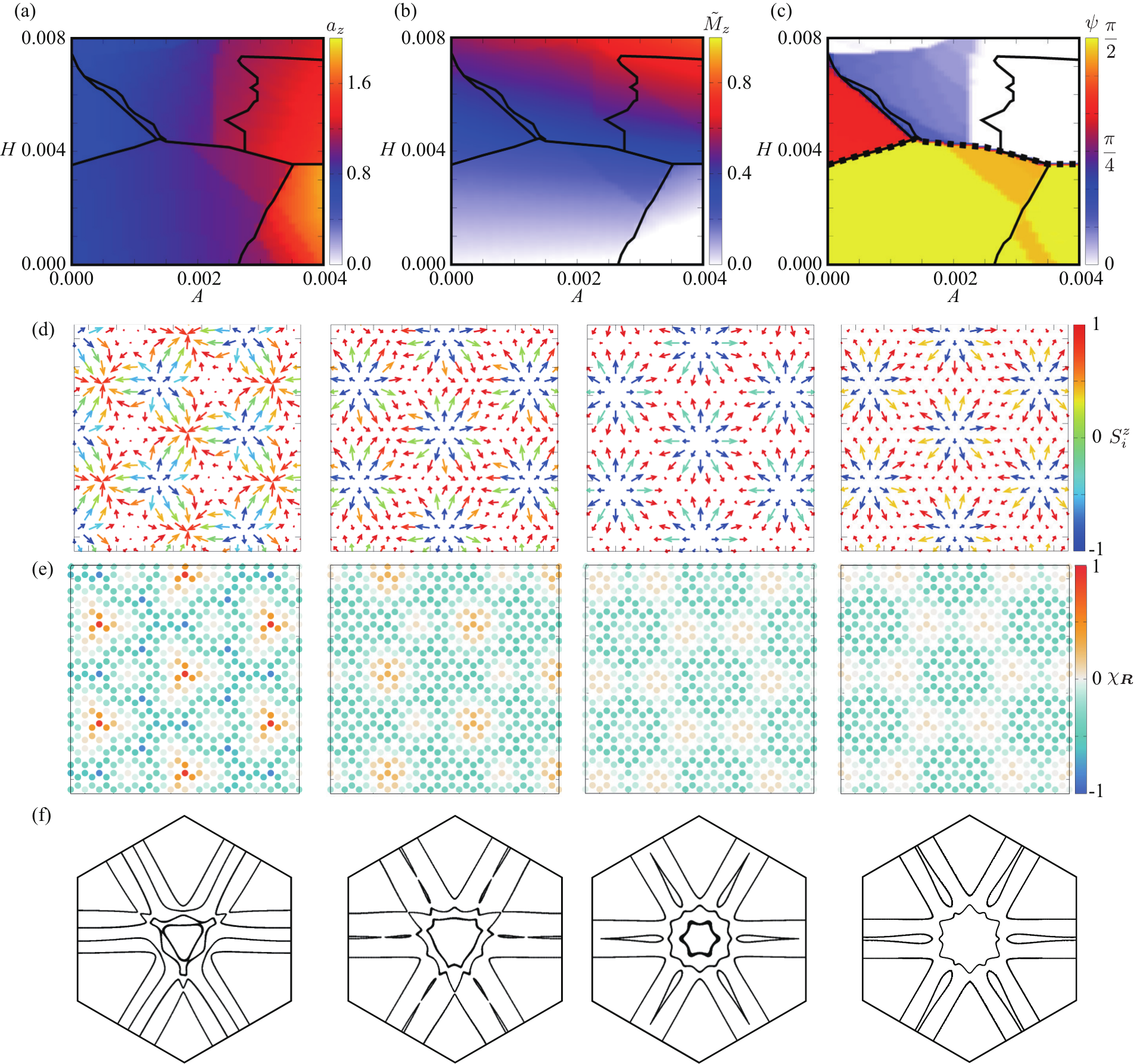} 
\caption{
\label{fig:paradep}
$A$ and $H$ dependences of the variational parameters (a) $a_z$, and (b) $\tilde{M}_z$, and (c) $\psi$ corresponding to Fig.~\ref{fig:PD12}. 
(d) The optimized spin configurations of (left) the pattern IV for $A=0$ and $H=0.005$, (middle left) the pattern II for $A=0.002$ and $H=0.005$, (middle right) the pattern I for $A=0.004$ and $H=0.005$, and (d) (right) the pattern II for $A=0.004$ and $H=0.008$. 
(e) The chirality configurations corresponding to (d). 
(f) The Fermi surfaces in the folded Brillouin zones for the magnetic unit cell corresponding to (d). 
}
\end{center}
\end{figure*}

Among the phase degrees of freedom, we especially focus on the position of the SkX core, which has not been discussed in previous literatures.
Nevertheless, the core position is important, as it is related to the symmetry of the SkXs. 
In the $|n_{\rm sk}|=1$ SkX, the offsite (onsite) core position breaks (preserves) the sixfold rotational symmetry of the triangular lattice [see Figs.~\ref{fig:PD}(a) and \ref{fig:PD}(b)].
Meanwhile, the offsite (onsite) core position in the $|n_{\rm sk}|=2$ SkX  breaks (preserves) the  space inversion and sixfold rotoreflection symmetries of the triangular lattice [see Figs.~\ref{fig:PD}(e) and \ref{fig:PD}(f)].  

For the above variational states with the patterns I-IV spin textures, we compare the grand potential at zero temperature, $\Omega= \langle \mathcal{H} \rangle/N $ ($N$ is the system size), and determine the lowest energy state. 
The variational parameter $\psi$ is taken at $\psi=\pi l/240$ ($l=0,1,2, \cdots, 120$) for the patterns I and II, while we take $\psi=\pi l/240$ ($l=0,1,2, \cdots, 240$) for the patterns III and IV.
The other variational parameters, $a_z$ and $\tilde{M}_z$, are taken at $a_z=0.02 l'$ ($l'=0,1,2, \cdots, 300$) and $\tilde{M}_z=0.002 l''$ ($l''=0,1,2, \cdots, 2500$). 
We consider these ordered states for the system size with $N=96^2$ and compute $\Omega$ by dividing the system into a 144-site unit cell and 64 supercells under the periodic boundary conditions. 

Figure~\ref{fig:PD12} shows the variational phase diagram of the double exchange model in Eq.~(\ref{eq:Ham}) while varying $A$ and $H$. 
The spin patterns I, II, and IV appear in the phase diagram, while the pattern III is not stabilized in the present parameter region. 
The dashed curve in Fig.~\ref{fig:PD12} stands for the boundary between the $|n_{\rm sk}|=1$ and $|n_{\rm sk}|=2$ SkXs. 
The corresponding variational parameters $a_z$, $\tilde{M}_z$, and $\psi$ are shown in Figs.~\ref{fig:paradep}(a), \ref{fig:paradep}(b), and \ref{fig:paradep}(c), respectively.
The behaviors of $a_z$ and $\tilde{M}_z$ in Figs.~\ref{fig:paradep}(a) and \ref{fig:paradep}(b) are intuitively understood from the energy gain in the presence of $A$ and $H$; $a_z$ tends to be larger for large $A$, since the easy-axis anisotropy aligns the spins along the $\pm z$ direction. 
Similarly, $\tilde{M}_z$ tends to be larger for large $H$, since the magnetic field aligns the spin along the $+ z$ direction.
Furthermore, we find that $\psi$ also changes depending on $H$ and $A$, as shown in Fig.~\ref{fig:paradep}(c). 
Here, the type of the spin density wave changes from the sinusoidal wave ($\psi=\pi/2$) for a low magnetic field to the spiral wave ($\psi=0$) for a high magnetic field.
This behavior is attributed to the $\psi$ dependence of the magnetization; the magnetization for $\psi=0$ tends to be larger than that for $\psi=\pi/2$ in the triple-$Q$ spin texture in Eq.~(\ref{eq:spin})~\cite{yambe2021skyrmion}. 

For the isotropic double exchange model at $A=0$, the $|n_{\rm sk}|=2$ SkX with the spin pattern II is stabilized for $0\leq H \lesssim 0.00325$, while the $|n_{\rm sk}|=1$ SkX with the spin pattern IV or II is stabilized for $0.00325 \lesssim H$. 
The critical field of the transition between SkXs with $|n_{\rm sk}|=1$ and $|n_{\rm sk}|=2$ is almost consistent with that obtained by the unbiased KPM-LD simulations~\cite{Ozawa_PhysRevLett.118.147205}. 
It is noted that the $|n_{\rm sk}|=1$ SkX for $0.0065 \lesssim H$ is metastable, since the anisotropic triple-$Q$ state which is not taken into account in the variational states in Eq.~(\ref{eq:spin}) is stabilized. 
The variational results show that the skyrmion cores in the isotropic double exchange model are locked at the interstitial sites, as shown in Figs.~\ref{fig:PD}(b), \ref{fig:PD}(d), and \ref{fig:PD}(f), which indicates sixfold rotational and inversion symmetries of the triangular lattice is broken in both the $|n_\mathrm{sk}|=1$ and $|n_\mathrm{sk}|=2$ SkXs.

Although the $|n_{\rm sk}|=2$ SkX in the pattern II region in the phase diagram has a similar spin texture to that in Fig.~\ref{fig:PD}(f), the $|n_{\rm sk}|=1$ SkX in the pattern IV region 
shows a different spin texture from that with $\psi=0$ in  Fig.~\ref{fig:PD}(d).
In this region, the $|n_{\rm sk}|=1$ SkX is characterized by $\psi \sim \pi/3$. 
In other words, the spin texture is described by the superposition of the sinusoidal wave and the spiral wave, which is different from the SkX represented by the superposition of three spirals in chiral/polar magnets [see Figs.~\ref{fig:PD}(a) and \ref{fig:PD}(b)].
The spin configuration of the pattern IV at $A=0$ and $H=0.005$ is shown in the left panel of Fig.~\ref{fig:paradep}(d), which consists of three types of vortices: one with vorticity $-2$ around $S^z \simeq +1$, another with vorticity $+1$ around $S^z \simeq +1$ and the other with vorticity $+1$ around $S^z \simeq -1$. 
The corresponding spin chirality distribution is shown in the left panel of Fig.~\ref{fig:paradep}(e), where the integration of the topological charge proportional to the spin chirality gives the quantized number of minus one ($n_{\rm sk}=-1$). 
Reflecting the breaking of sixfold rotational symmetry, the Fermi surface in the ordered state is threefold rotational symmetric, as shown in the left panel of Fig.~\ref{fig:paradep}(f).

By introducing $A$, the optimal spin configuration of the $|n_{\rm sk}|=1$ SkX in the pattern IV region is mainly replaced by the pattern II spin configuration while keeping $|n_{\rm sk}|=1$ except for the narrow pattern I region sandwiched between the patterns IV and II regions. 
The spin texture in the pattern II is characterized by small $\psi$, as shown in Fig.~\ref{fig:paradep}(c), whose spin and chirality configurations are shown in real space in the middle left panels of Figs.~\ref{fig:paradep}(d) and \ref{fig:paradep}(e), respectively.
Compared to the left and middle left panels in Figs.~\ref{fig:paradep}(d) and \ref{fig:paradep}(e), one can find that both the spin and scalar chirality textures seem to be similar with each other.
Thus, it is difficult to distinguish the patterns II and IV in the external magnetic field. 
Meanwhile, the important observation is that the skyrmion core is still placed at the interstitial site when introducing $A$, although the pattern I appears in the narrow region between the patterns II and IV.  
As the skyrmion core is located at the interstitial site, the threefold-symmetric Fermi surface is also obtained in the pattern II, as shown in the middle left panel of Fig.~\ref{fig:paradep}(f).

While further increasing $A$, the pattern II state turns into the pattern I state for both the $|n_{\rm sk}|=1$ and $|n_{\rm sk}|=2$ SkXs, as shown in Fig.~\ref{fig:PD12}.   
In the region for the $|n_{\rm sk}|=1$ SkX, the spin texture is characterized by $\psi=0$, which indicates that the individual spin density wave is described by the spiral state. 
In this case, the skyrmion core is lied at the onsite position, as shown in the middle right panel of Fig.~\ref{fig:paradep}(d). 
Accordingly, the scalar chirality is also distributed in a sixfold-symmetric way, as shown in the middle right panel of Fig.~\ref{fig:paradep}(e). 
In the end, the Fermi surface is sixfold symmetric in the middle right panel of Fig.~\ref{fig:paradep}(f)~\footnote{The momentum-dependent in-plane spin polarization occurs so that inversion symmetry is broken while keeping sixfold rotational symmetry.}. 
This sixfold Fermi surface is again broken by increasing the magnetic field $H$, where the pattern I state changes into the pattern II one. 
The skyrmion core in the pattern II state is located at the interstitial sites as shown in the right panels of Figs.~\ref{fig:paradep}(d) and \ref{fig:paradep}(e), and then the sixfold rotational symmetry of the Fermi surface is slightly broken in the right panel of Fig.~\ref{fig:paradep}(f). 

For low-field $H$, the optimal spin configuration of the $|n_{\rm sk}|=2$ SkX in the pattern II region is almost the same against $A$ up to $A \lesssim 0.0026$, as shown in Fig.~\ref{fig:PD12}. 
Similar to the $|n_{\rm sk}|=1$ SkX, the skyrmion core position moves to the onsite position for large $A$, which is denoted as the pattern I. 
However, the spin texture in the $|n_{\rm sk}|=2$ SkX in Eq.~(\ref{eq:spin}) has higher energy than the other spin ansatz in the form of $\bm{S}_i \propto (\cos \mathcal{Q}_{1i}, \cos \mathcal{Q}_{2i}, a_z\cos \mathcal{Q}_{3i})$ with $|n_{\rm sk}|=2$ when the unbiased KPM-LD simulations are performed~\cite{Hayami_PhysRevB.99.094420}. 
This is distinct from the situation where the $|n_{\rm sk}|=1$ SkX in Eq.~(\ref{eq:spin}) is stabilized under the magnetic field, as discussed above; indeed, we show that the $|n_{\rm sk}|=1$ SkX expected from the variational calculations remains stable in the presence of $A$ even after carrying out the KPM-LD simulations, as will be discussed in the next section~\cite{Hayami_PhysRevB.99.094420}.

The above results indicate that the skyrmion cores are located at the interstitial sites without the anisotropy, while they move to the onsite when increasing $A$. 
This behavior is reasonable because the easy-axis anisotropy favors the large $z$-spin component, whose energy gain is maximized when the skyrmion core with $S_i^z \simeq -1$ is placed at the site. 
Although it is usually difficult to control $A$ by external stimuli, one can also change the position of the skyrmion core by applying the external magnetic field for materials with the large easy-axis anisotropy. 
The change of the core position can be directly detected in $\bm{k}$-resolved angle-resolved photoemission spectroscopy as a consequence of different symmetry of the Fermi surface.
It is noteworthy such information to identify complicated spin textures by an electric probe has been achieved by the noncollinear magnetoresistance~\cite{hanneken2015electrical,Kubetzka_PhysRevB.95.104433} and the spectroscopic imaging scanning tunneling microscopy measurement~\cite{Yasui2020imaging}, since the formation of the (multiple) spin density waves affects local electronic states and causes charge density waves~\cite{Hayami_PhysRevB.104.144404}. 
Our results indicate that angle-resolved photoemission spectroscopy becomes another electric probe of the SkX spin textures in a complement way. 
Moreover, as the antisymmetric band structure with respect to $\bm{k}$ is related with the nonreciprocal transport even without the spin-orbit coupling~\cite{Hayami_PhysRevB.90.024432,Ishizuka2020anomalous,Hayami_PhysRevB.101.220403,Hayami_PhysRevB.102.144441}, our results indicate that the control of the transport properties is possible in the SkXs in itinerant magnets.

\section{Langevin dynamics simulation}
\label{sec:Langevin dynamics simulation}

\begin{figure}[t!]
\begin{center}
\includegraphics[width=1.0\hsize]{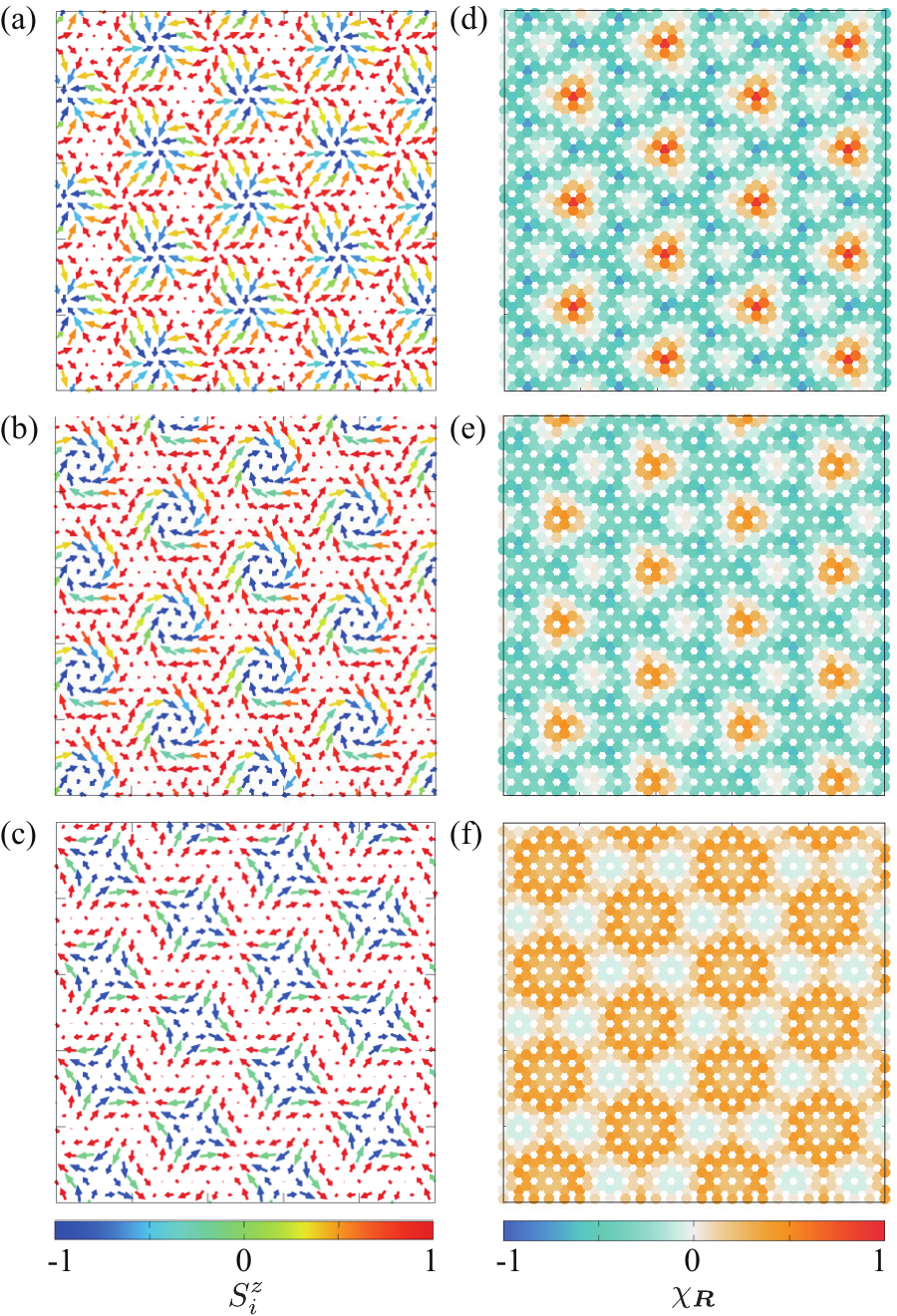} 
\caption{
\label{fig:KPM}
(a), (b), (d), (e) The results obtained by the KPM-LD simulations. 
The spin configurations for (a) $A=0$ and $H=0.006$ and (b) $A=0.002$ and $H=0.005$. 
(d) and (e) stand for the scalar chirality configurations corresponding to (a) and (b), respectively. 
(c), (f) The spin [(c)] and scalar chirality [(f)] configurations obtained by the simulated annealing for the effective spin model at $K/\tilde{J}=0.5$, $A/\tilde{J}=0.2$, and $H/\tilde{J}=0.4$. 
}
\end{center}
\end{figure}

To investigate the stability tendency of the $|n_{\rm sk}|=1$ SkX in the previous section, we perform the KPM-LD simulation, which is an unbiased numerical simulation based on Langevin dynamics~\cite{Barros_PhysRevB.88.235101} with the kernel polynomial method~\cite{Weis_RevModPhys.78.275}. 
As this method has an advantage of performing a large-scale calculation compared to the standard Monte Carlo simulation combined with the direct diagonalization, it has been applied to similar models consisting of free fermions and one-body boson field~\cite{Barros_PhysRevB.88.235101,Barros_PhysRevB.90.245119,Ozawa_doi:10.7566/JPSJ.85.103703,Ozawa_PhysRevLett.118.147205,Wang_PhysRevLett.117.206601,Ozawa_PhysRevB.96.094417,Chern_PhysRevB.97.035120,ono2019photoinduced,Eto_PhysRevB.104.104425,Hayami_10.1088/1367-2630/ac3683}. 
The simulation is performed at zero temperature for the system size with $N=96^2$ under the periodic boundary conditions. 
In the kernel polynomial method, we expand the density of states by up to 2000th order of Chebyshev polynomials with $16^2$ random vectors~\cite{Tang12}. 
In the Langevin dynamics, we use a projected Heun scheme~\cite{Mentink10} for 1000-3000 steps with the time interval  $\Delta \tau =2$. 

Figure~\ref{fig:KPM}(a) shows the real-space spin configuration at $A=0$ and $H=0.006$ obtained by the KPM-LD simulations. 
The spin texture well corresponds to that obtained by the variational calculations in the left panel of Fig.~\ref{fig:paradep}(d): it consists of three types of vortices, one with vorticity $-2$ around $S^z \simeq +1$, another with vorticity $+1$ around $S^z \simeq +1$ and the other with vorticity $+1$ around $S^z \simeq -1$. 
We also reproduce the spin configuration of the pattern II in the middle left panel of Fig.~\ref{fig:paradep}(d) by performing the KPM-LD simulations at $A=0.002$ and $H=0.005$ in Fig.~\ref{fig:KPM}(b). 
The scalar chiralities in real space in Figs.~\ref{fig:KPM}(d) and \ref{fig:KPM}(e) correspond to those in left and middle left panels of Fig.~\ref{fig:paradep}(e) as well. 
Similarly, the $n_{\rm sk}=2$ SkX at $A=0$ is obtained by the KPM-LD simulations~\cite{Ozawa_PhysRevLett.118.147205}.

However, we could not obtain the spin texture corresponding to that in the middle right panel of Fig.~\ref{fig:paradep}(d) in the KPM-LD simulations, since the simulations are easily trapped into the metastable state with the multi-domain configurations in terms of the spin and chirality, which makes the core position ambiguity.
This is presumably attributed to the small energy scale under the large single-ion anisotropy. 
Instead of that, we analyze an effective spin model derived from the double exchange model, where the first four terms in Eq.~(\ref{eq:Ham}) are replaced by the bilinear RKKY interaction $\tilde{J}$ and the positive biquadratic interaction $K$ (see Ref.~\onlinecite{Hayami_PhysRevB.95.224424} for the model in details). 
By taking the model parameters as $K/\tilde{J}=0.5$, $A/\tilde{J}=0.2$, and $H/\tilde{J}=0.4$, adopting the same ordering vectors, and performing the simulated annealing following the manner in Ref.~\onlinecite{Hayami_PhysRevB.95.224424}, we confirmed that the skyrmion core is located at the site for a relatively large $A$, as found in the middle right panel of Fig.~\ref{fig:paradep}(d). 
The real-space spin and scalar chirality configurations obtained by the simulated annealing are shown in Figs.~\ref{fig:KPM}(c) and \ref{fig:KPM}(f), respectively.

\section{Summary}
\label{sec:Summary}

To summarize, we have investigated the optimal spin configurations of the SkXs in itinerant magnets by taking into account the phase degrees of freedom among the constituent waves. 
We find that the core positions of the SkX consisting of the commensurate ordering wave vectors with the crystal lattice are placed at the interstitial sites for small magnetic anisotropy, while those are found at the sites for large magnetic anisotropy by the variational calculations for the double exchange model on the discrete triangular lattice. 
We also find that the SkX with $|n_{\rm sk}|=1$ consists of the superposition of the triple sinusoidal and spiral waves for small magnetic anisotropy, while that is described by the triple spiral waves for large magnetic anisotropy. 
We discuss the possibility of the control of the nonreciprocal transport by applying the magnetic field (or changing the single-ion anisotropy), since the core positions at the interstitial site (onsite) break (hold) the sixfold rotational symmetry, leading to the threefold(sixfold)-symmetric band deformation. 
On the basis of the different Fermi-surface geometry resulting from the different core position, we propose that angle-resolved photoemission spectroscopy can be a useful electric probe to obtain information about the real-space spin textures.
Finally, we confirm that the spin textures expected from the variational calculations are obtained by the unbiased KPM-LD simulations for the double exchange model or simulated annealing for the effective spin model.

The present results provide information to examine the detailed symmetry of not only the SkXs and but also the hedgehog lattices with the commensurate magnetic modulations where the discreteness of the lattice structure is important~\cite{hayami2020phase,Shimizu_PhysRevB.103.054427}. 
Especially, the SkXs and the hedgehog lattices stabilized by the spin-charge entanglement in itinerant magnets might be promising, since the commensurate locking often happens owing to the local gap formation of the electronic band structure, as found in the chiral soliton lattice~\cite{matsumura2017chiral,okumura2018lock}. 
In addition, the results would be useful to investigate the effect of the intrinsic lattice pinning in the current-driven dynamics~\cite{Xia_PhysRevApplied.11.044046,zhang2020skyrmion}.

\begin{acknowledgments}
S.H. acknowledges Y. Motome for enlightening discussions in the early stage of this study.
This research was supported by JSPS KAKENHI Grants Numbers JP19K03752, JP19H01834, JP21H01037, and by JST PRESTO (JPMJPR20L8).
R.Y. was supported by Forefront Physics and Mathematics Program to Drive Transformation (FoPM).
Parts of the numerical calculations were performed in the supercomputing systems in ISSP, the University of Tokyo.
\end{acknowledgments}

\bibliographystyle{apsrev}
\bibliography{ref}

\end{document}